# Ultra-fast Vapor Generation by a Graphene Nano-ratchet

Hongru Ding[1,2,#], Guilong Peng[1,2,#], Dengke Ma[1,2], S.W. Sharshir[1,2] and Nuo Yang[1,2,*]

[1]State Key Laboratory of Coal Combustion, Huazhong University of Science and Technology (HUST), Wuhan 430074, P. R. China

[2]Nano Interface Center for Energy (NICE), School of Energy and Power Engineering, Huazhong University of Science and Technology (HUST), Wuhan 430074, P. R. China

[#] H. D. and G. P. contributed equally to this work.

Electronic mail: N.Y. (nuo@hust.edu.cn)

# ABSTRACT

Vapor generation is of prime importance for a broad range of applications: domestic water heating, desalination and wastewater treatment, etc. However, the natural evaporation is slow and low efficiency. Ratchet effect can give rise to nonzero mass flux under a zero-mean time-dependent drive. In this paper, we proposed a nano-ratchet, multilayer graphene with cone-shaped nanopores (MGCN), to accelerate the vapor generation. By performing molecular dynamics simulations, we found that the air molecules spontaneously transport across MGCN and form a remarkable pressure difference between the two sides of MGCN. Besides, we studied the dependence of pressure difference on the ambient temperature and the geometry of MGCN in detail. By further analysis of the diffusive transport, we identified that the pressure difference relates to the competition between ratchet transport and Knudsen diffusion. The pressure difference could give rise to 15 times enhancement of vapor generation at least, which shows there is a widely potential application of the ratchet effect of MGCN.

# Introduction

Water vapor generation has a broad range of applications, from power generation[1, 2], desalination[3], water purification[4] to oil recovery[5] and so on. Whereas, the limited evaporation rate impeded the further development of those field. Over the past decades, many researches including localizing heat, using nanoparticles or thin-film and constructing vacuum have been carried out to improve vapor generation.[6-13] Among them, providing vacuum has proved very effective: In 2007, Nassar et al.[12] achieved 303% improvement of the productivity of desalinized water under vacuum of 25 kPa by using a vacuum pump; and later, Kabeel et al.[13] obtained an increase of 53.2% in the total daily distillation when provided vacuum with a fan. However, all of the traditional methods to create vacuum need extra energy and devices, which increases the cost and complexity of the system. Obviously, a simple and environment friendly way to create vacuum for vapor generation is in pressing need.e

Interestingly, we found that the ratchet effect[14] might provide a brand new approach to get vacuum. The ratchet theory, found by Smoluchowski in 1912[15], indicates that asymmetric potentials, such as thermal gradient[16, 17], magnetic gradient[18], electrochemical gradient[19, 20] and electric field[21, 22], are able to produce net mass or energy flux under zero-average dynamic load. Asymmetric structures acting as the ratchet were also reported in recent years.[23-26] Additionally, Doering et al.[27-29] have demonstrated the existence of noise-driven ratchet, which consumes energy extracted from thermal fluctuations of the environment. Therefore, unidirectional air molecules flux can be introduced by asymmetric nanostructures, through rectifying thermal fluctuations.

Besides，both theoretical studies[30-32] and experiments[33, 34] have proved carbon nanotubes (CNT) as fast water transporters for its regular structure and the hydrophobic nature.[35] With the same hydrophobic nature and similar structure, multilayer graphene with cone-shaped nanochannel is expected to keep the fast water molecules flow rate and this will guarantee the significant reduction of energy intensity.

Here, we propose a carbon nano-rachet, the multilayer graphene with cone-shaped nanopores (MGCN), which can spontaneously pump air molecules from the low pressure side to the high. We also investigated the following factors that can affect the pumping efficiency: the number of graphene layers, the aperture of the cone, the cone angle and the ambient temperature. Moreover, we suggest the physical mechanism of the MGCN's ratchet effect and discuss its potential to be applied in vapor generation. This may provide a new way for studying the molecular transport in asymmetric membranes and promote the vapor generation's development.

## Method

The simulation domain consists of two blocks of saturated moist air separated by MGCN (Fig. 1). MGCN is constructed by $N_l$ layers graphene with cone-shaped nanopores, where the area of the minimum pore is $A_{min}$ and the cone angle is α. The periodic boundary conditions were applied in X and Y directions and rigid walls were applied in the ends of Z direction. Cone-shaped nanopores were created by removing carbon atoms within different distance from the center of a hexagonal graphene ring. Initially, the air pressure ($P_0$) of the two identical blocks is 101 kPa and the molecular ratio between $N_2$, $O_2$ and $H_2O$ molecules is 26:8:1. Figure 1 also vividly illustrates the physics of the ratchet effect in MGCN. The thermal fluctuations act as the rod, running back and forth. The asymmetric potential acts as the pawl, preventing the anticlockwise rotation. Then the gear rotates clockwise; in other words, directed transport of air are obtained.

Simulations were performed with the LAMMPS package[36], with a time step of 1 fs at $T_{amb}$ K. The intermolecular interactions, essential to this study, were modeled by Lennard-Jones(LJ) potential $V(r) = 4\epsilon[(\frac{\sigma}{r})^{12} - (\frac{\sigma}{r})^{6}]$ plus Coulomb potential. Two centered LJ potential models[37] and rigid TIP4P[38] model are used for the $N_2$ and $O_2$ pairwise potential and water molecules, respectively. Long-range electrostatic forces were computed with the P³M method. For graphene, the LJ parameters were adopted from the work of Beu et al[39, 40]. For all pairwise LJ terms, the Lorentz-Berthelot mixing rules were applied and the cutoff distance in LJ potential was set to 2.5σ.

NEMD simulations in the NVT ensemble were used to compute the density distribution and pressure of the air, after the relaxation of the system in the NVT ensemble. Nosé−Hoover thermostats[41] were applied to the air molecules. To obtain meaningful statistics, for each set of parameters, more than 8 independent simulations were performed sufficiently long (over 8 ns). In all simulations, to substantially reduce the computational cost, all the graphene atoms were assumed to be held rigid. Further simulation details can be found in Supporting Information (SI). What's more, the time step and size effects on the ratchet transport herein are also discussed in SI.

## Results and Discussions

The main results are shown in Fig. 2. We tracked the density of air in all regions over time. A typical air's density profile over time is shown in Fig. 2(a). This simulation is performed at 300 K. $N_l$, $A_{min}$ and α are equal to 4, 55.2 Å² and arctan0.25, respectively. As shown in Fig. 2(a), it is worth noting that there is a significant density difference between the left and right regions at 300 K, ∼ 0.2 g/m³. This unexpected phenomenon results from the ratchet effect, producing net mass transport by rectifying thermal fluctuations. For the sake of further enhancing the evaporation (ratchet) efficiency, we investigated the relationship between the pressure difference ΔP with the number of graphene layers, the area of the minimum cone-shaped pore, the cone angle and the ambient temperature.

Firstly, the dependence of ΔP on $N_l$ was studied, when $A_{min}$ and α were fixed at 33.1 Å² and arctan0.25, respectively. As shown in Fig. 2(b), when the $N_l$ increases from 2 to 5, ΔP decreases linearly. To obtain the in-depth comprehension of this dependence, the force distribution (along Z axis) around MGCN was calculated by using a nitrogen atom as the probe and shown in Fig. 3. The color bar and arrows describe the magnitudes and directions of the force. Vertical axis and Horizontal axis correspond to the Y and Z axis, respectively. For better comprehension, we plot the graphene layers in this figure. A distinct symmetry breaking between the positive (rightward) and negative (leftward) force appears in the cone-shaped nanopores. According to the force distribution of MGCN with different $N_l$, it's clear that the negative force (blue regions) becomes weaker but the positive force (red regions) remains nearly the same in the nanochannel. That's why the ratchet effect weakens with the increasing of layers.

Apart from $N_l$, another important factor, $A_{min}$ was also discussed here. In this part, all simulations were performed at room temperature with fixed $N_l$ (4) and α (arctan0.25). The results are shown in Fig. 2(c). As the $A_{min}$ changes from 22.1 Å² to 55.2 Å², ΔP doubles, ~ 16 kPa; yet ΔP sharply reduces to 3.3 kPa when $A_{min}$ increases to 77.3 Å². The increasing tendency is seemingly consistent with the variation of force distribution along Z axis in the nanochannel (Fig. 4(b-d)), where the negative (leftward) force regions expand and the positive (rightward) force regions shrink significantly with the increasing $A_{min}$.

However, from 22.1 Å² to 77.3 Å², the negative force weakens gradually. It is detrimental to the ratchet transport and make things unclear. Therefore, we need to seek more interpretation by analyzing the potential distribution in the nanochannel, which is an indispensable factor for the ratchet effect. The potential of MGCN with different $A_{min}$ are shown in Fig. 4 (a). Vertical axis and Horizontal axis correspond to the potential values and Z axis, respectively. The potential peaks, i.e., the energy barriers, locate at the position of each graphene layer and hinder the transport of atoms. Obviously, from left to right, the potential peaks become lower and lower, which means the molecules in right region get a higher probability to enter the nanochannel. Such asymmetry results in atoms flowing more easily along the specific direction. When $A_{min}$ increases, the number and value of peaks become less and smaller. Especially when $A_{min}$ reaches to 55.2 Å$^2$, the two rightmost peaks even disappear. Consequently, compared with quantitative changes of negative force, the broader nanochannel and the less and lower energy barriers are dominant, hence the strengthened ratchet effect. As for the reasons of the decreasing trend, which involves the diffusion transport (rightward), will be explained later. Similarly, ΔP also has nonmonotonic dependence on the truncated-cone angle α, shown in Fig. S4. That is why the adopted tanα of other simulations is fixed at 0.25 in this paper.

Herein, the ambient temperature is also a crucial factor, which decides the force of zero-average dynamic load, i.e., thermal fluctuations. Because thermal fluctuations can not be considered as white noise with negligible time correlation, for the correlation lengths of them becoming significantly long for nanopores[39]. If we attempt to advance the ratchet, there will be a minimum force necessary to overcome the barriers. That is, the ambient temperature should be high enough to activate the ratchet. However, if the temperature is too high (the force is too powerful), the ratchet could also run in the opposite direction. Because, compared with the power of thermal fluctuations, the asymmetric potential is too small to impose the rightward movements of the air any more. Then the ratchet effect is broken. We performed the simulations with fixed $N_l$ (4), $A_{min}$ (55.2 Å$^2$) and tanα (0.25) at different temperature, and results are shown in Fig. S3. At the range of 300 K to 600 K, the resulting ΔP are 20.1 kPa and 9.2 kPa, and this difference is consistent with our speculation.

The previous discussions of results mainly focus on the strength of the ratchet effect. However, the diffusive transport, resulted from the concentration difference, limits the further growth of concentration. Therefore, the final distribution of air depends on the competition between the ratchet transport (leftward) and diffusive transport (rightward). Since the scale length of the truncated-cone nanopores is much smaller than the mean free path of the air molecules, the Knudsen diffusion[42] occurs here. The Knudsen diffusion flux is defined as (details are shown SI),

$$\Phi_k = -\frac{2r_{min}+(N_l-1)h\tan\alpha}{3}u\frac{dC}{dz} \tag{1}$$

where $r_{min}$ is radius of the minimum pores; $u$ is the characteristic velocity of air molecules; $\frac{dC}{dz}$ is the concentration gradient and $h$ is the interlayer spacing of MGCN.

When the number of MGCN layers increases from 2 to 5, $r_{min}$, $h$ and $\alpha$ remain the same. And the concentration of air has a negligible change, $\frac{dC}{dz}$ is assumed to be proportional to the thickness of MGCN, $(N_l - 1)h$. It's clear that, when $N_l$ increases, $\Phi_k$ only has subtle reduction, resulting from the product of $r_{min}$ and $\frac{dC}{dz}$. As the consequence of the significant weakening in ratchet effect and the subtle reduction of diffusion flux, ΔP surely decreases. As for the situation of increasing $A_{min}$, $r_{min}$ becomes bigger and other parameters remain the same, hence $\Phi_k$ also increases. However, the broadening nanochannel and the decreasing of energy barriers dominate. But when $A_{min}$ increases to 77.3 Å², the dominant positions are taken by the sustainable weakening of leftward force and strengthening of diffusive transport. Thus ΔP shows an increased and then decreased tendency.

To illustrate its potential in practical application, the effect of MGCN on evaporation is discussed in this paragraph. The previous results and discussions provide useful guidelines for the design of MGCN. According to the results in Fig. 2, MGCN can create a biggest pressure difference as 21 kPa, between the left and right region. Hence, when MGCN is used to decrease the vapor pressure near the water-vapor interface, the evaporation rate will be improved. The enhancement of evaporation $\eta_i$ can be described as (details are shown in SI):

$$\eta_i = \left(\frac{\varepsilon \cdot \Delta P \cdot P_S}{P_{atm}\dot{m}}\sqrt{\frac{M}{2\pi RT}} - 1\right) \times 100\% \tag{2}$$

where $P_{atm}$ is the atmospheric pressure, $\Delta P$ is the pressure difference between the two sides of MGCN. It should be noted that $\dot{m}$ is the evaporation flux without MGCN.

Thus, to calculate the enhancement, $\varepsilon$, T, and $\dot{m}$ of a traditional evaporation condition (i.e., without MGCN) should be measured.

Due to the lack of consensus in the value of $\varepsilon$, herein we regard it as an independent variable which varies from 0.001 to 1 [41-43]. Figure 5 shows the enhancement according to the temperature and evaporation data from references [3, 6, 43]. As we can see, even if $\varepsilon$ only equals to 0.001, an enhancement of more than 15 times can be expected. If $\varepsilon$ equals to 0.1, the enhancement will be thousandfold. The ultra-high evaporation rate at room temperature indicates that, by using MGCN, a lot of low grade energy can be used for vapor generation, such as low intensity solar energy or waste heat. Meanwhile, the low evaporation temperature can decrease the energy dissipation effectively. Those merits imply a high potential of practical application of MGCN.

## Conclusion

Summarizing, based on the ratchet effect, the MGCN we proposed can pump molecules unidirectional to enhance vapor generation by rectifying thermal fluctuations. By performing MD simulations, we find that MGCN can produce pressure difference as high as 21.0 kPa. It corresponds to more than 15 times enhancement compared with natural evaporation. We provide detailed discussions about the factors that may affect the efficiency of this nano-pump and provide guidelines for the design of MGCN. It is found that the pressure difference $\Delta P$ decreases with increasing $N_l$, but firstly increases and then decreases with increasing $A_{min}$ and $\tan\alpha$. Moreover, we find $\Delta P$ is also sensitive to the ambient temperature. According to the results, it is better to construct MGCN by using 2 layers graphene with cone-shaped nanopores, where the area of the minimum pore is 55.2 Å$^2$ and the cone angle is fixed as $\arctan 0.25$, then the best performance can be obtained. After comparatively analyzing the force and potential profile of MGCN in different conditions, we explained that mechanism of the ratchet transport and the dependence of $\Delta P$ on those factors. Through the analysis from diffusive transport, we further demonstrated that the final distribution of air is decided by the competition between ratchet transport and diffusive transport. The results suggest that enhancement of vapor generation can be achieved by the novel multilayer graphene environment friendly, which is much easier to fabricate and consumes only low grade energy like low intensity solar energy and waste heat.

## Acknowledgments

The project was supported by the National Natural Science Foundation of China 51576076 and 51711540031 (NY). The authors thank the National Supercomputing Center in Tianjin (NSCC-TJ) for providing help in computations. The authors are grateful to Zelin Jin, Quanwen Liao for useful discussions and Yingying Zhang for polishing of the manuscript.

## Author contributions

H.D., G.P. and N.Y. conceived the work. H.D. performed the numerical simulations. All the authors discussed and interpreted the data. H.D. and G.P. wrote the paper and all the authors commented on the contents of the manuscript. N.Y. directed the project.

## Additional information

Supplementary information is available in the online version of the paper.

## Competing financial interests

The authors declare no competing financial interests.

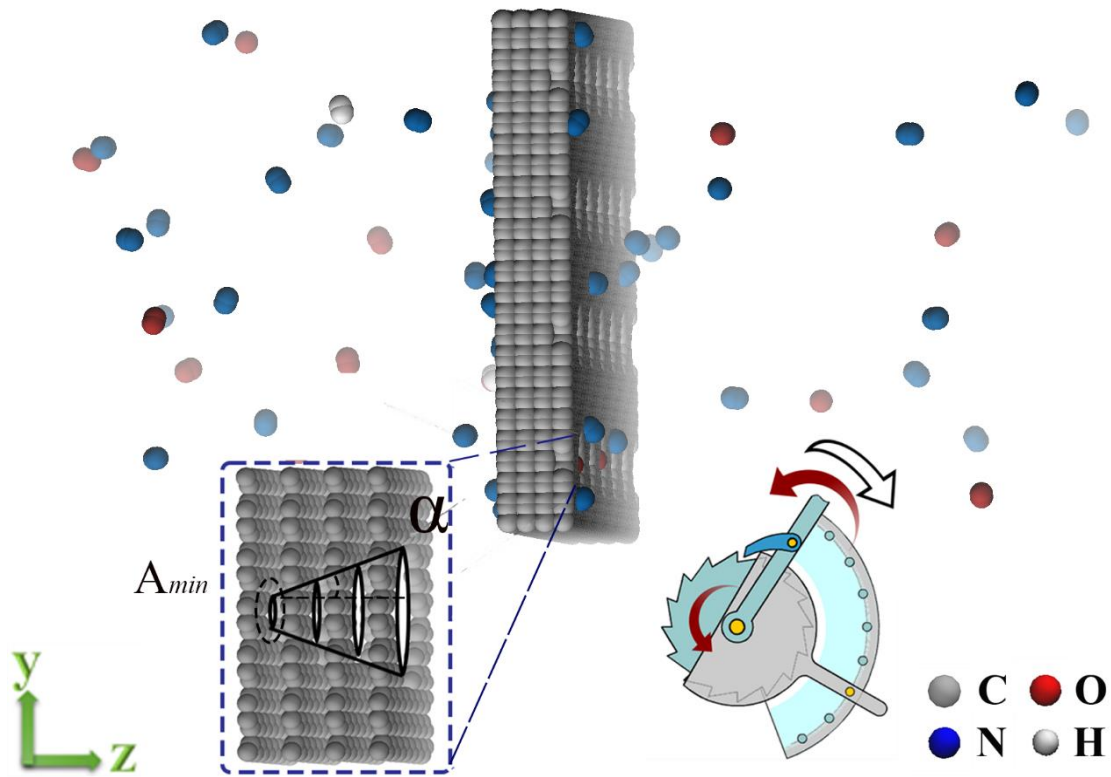

Figure 1. Schematic diagram for the simulation setup. Multilayer graphene membranes with cone-shaped nanopores are in grey; Nitrogen, oxygen and hydrogen molecules are in blue, red and white, respectively. Left inset: Magnification of the nanopores with minimum area $A_{min}$ and cone angle α. Right inset: The thermal fluctuations act as the rod, running back and forth. The MGCN acts as the pawl and rectify the work of the rod. Then the rotary direction of the gear is clockwise. That is the transport direction of the air atoms is leftward. The two arrows around the rod shows the directions of thermal fluctuations' work. And the arrow around the gear shows the direction of the ratchet transport.

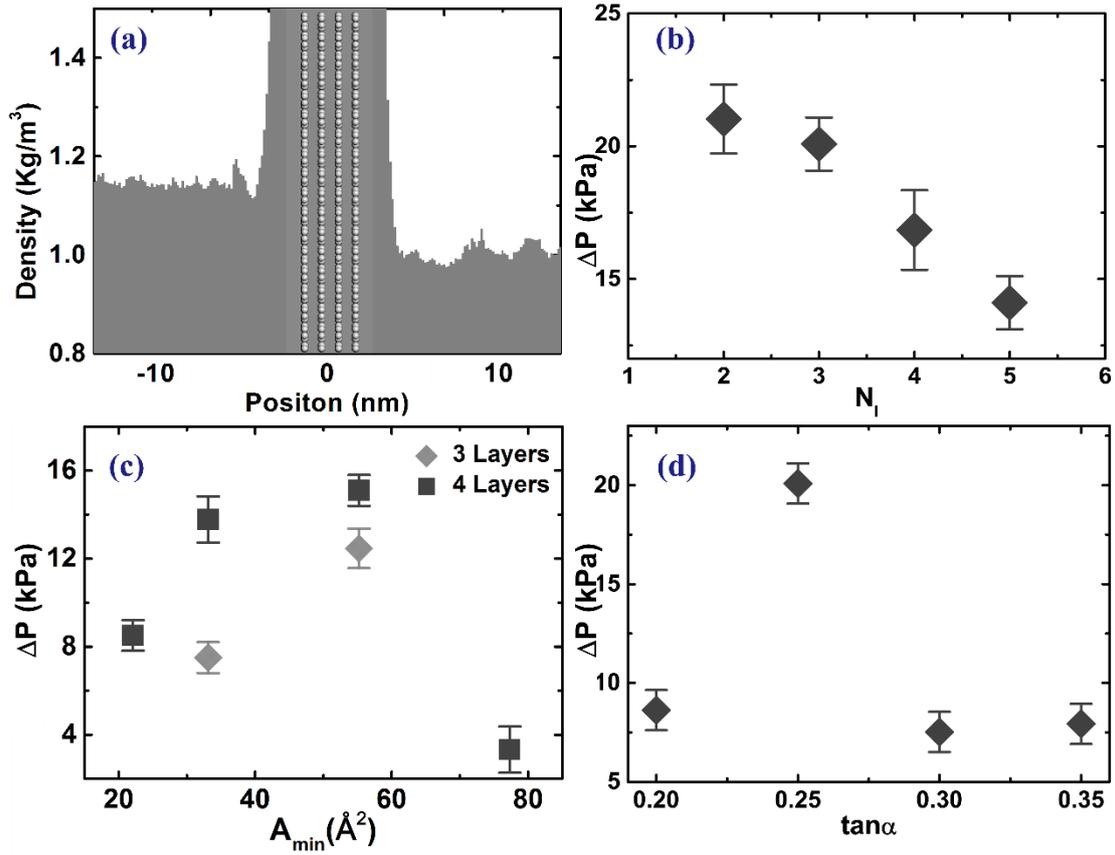

Figure 2. (a) A typical density profile of saturated moist air molecules. In this simulation, MGCN is constructed by 4 layers graphene with truncated cone nanopores, where the area of minimum pore is 55.2 Å$^2$ and the cone angle is arctan0.25. The dependence of the pressure difference ΔP on the number of graphene layers $N_l$ (b), the minimum pore's area $A_{min}$ (c) and the cone angle α (d).

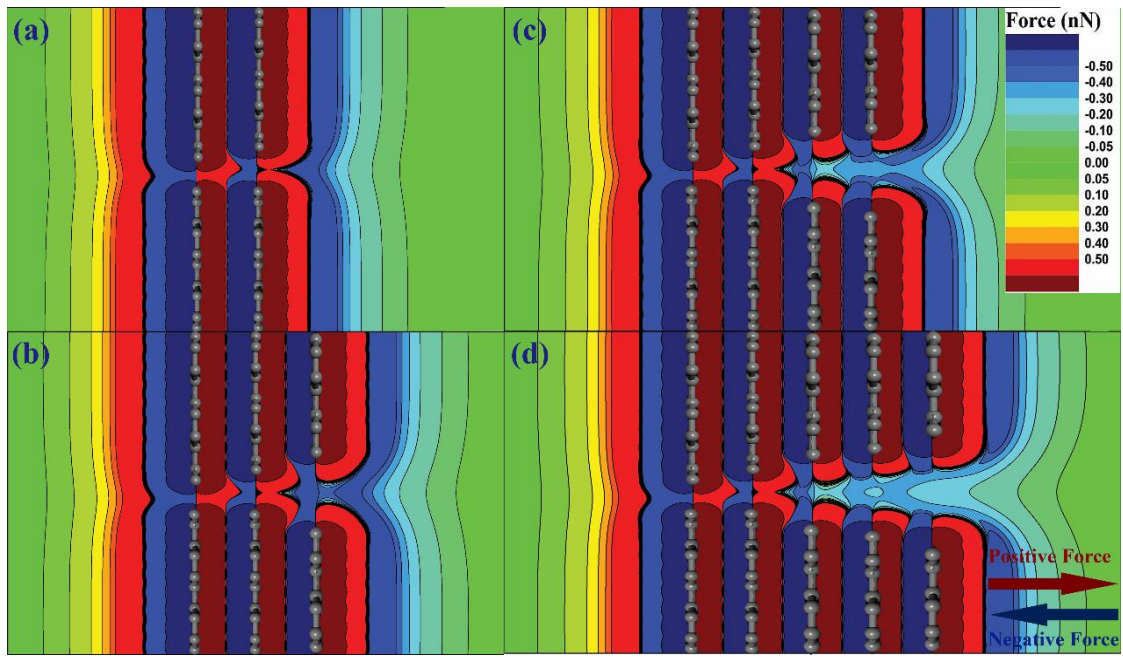

Figure 3. The force (along the Z direction) distribution of MGCN with different graphene layers (from 2 layers to 5 layers). The area of minimum pore is fixed as 33.1 Å² and the cone angle is arctan0.25. The arrows point out the direction of positive and negative forces.

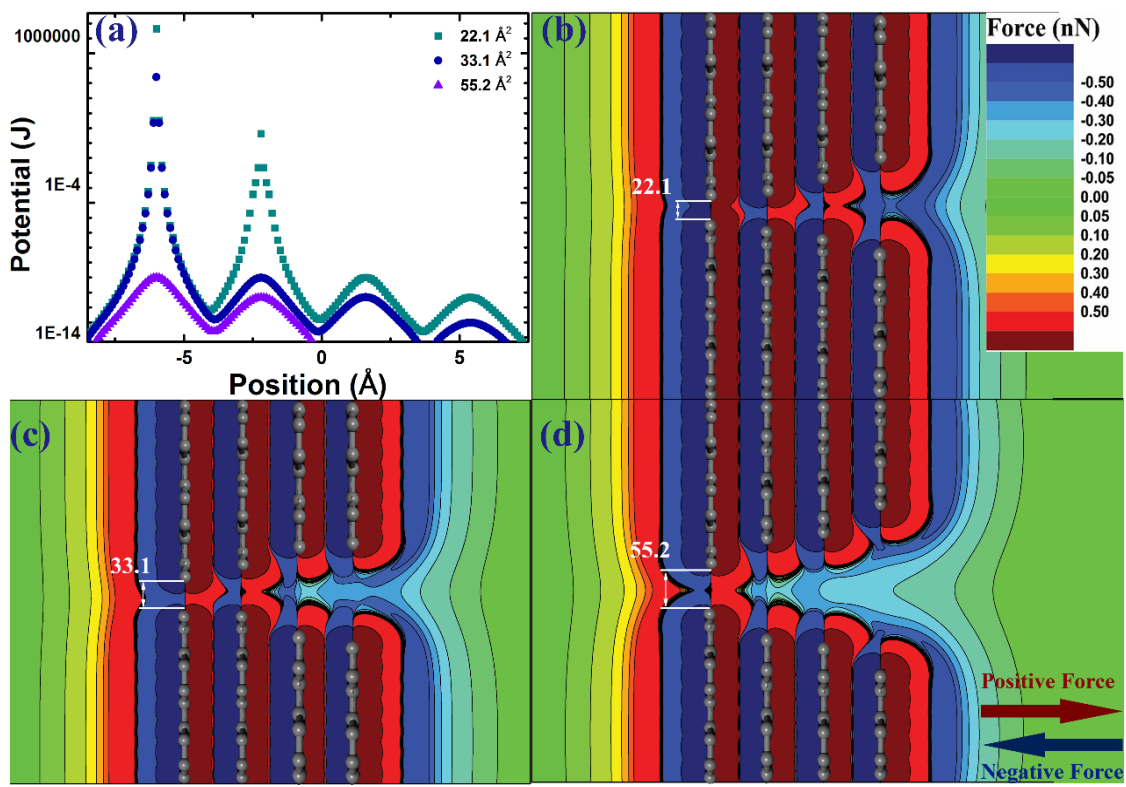

Figure 4. The potential (a) and the force (along the Z direction) (b)-(c) profile of the MGCN with different $A_{min}$ (from 22.1 Å² to 55.2 Å²) are shown in. The number of graphene layers is fixed as 4 and the cone angle is arctan0.25. The arrows point out the direction of positive and negative force.

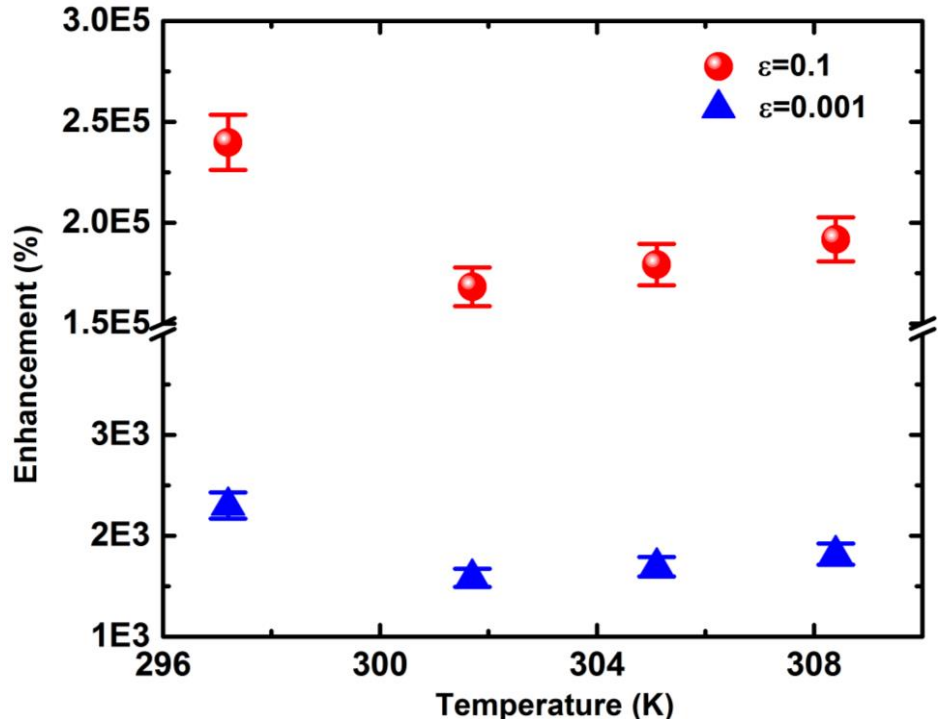

Figure 5. The enhancement of evaporation at different water temperature. For MGCN, the number of graphene layers, the cone angle, the area of minimum pore are fixed as 2, arctan0.25 and 33.1 Å², respectively.

Supporting Information

# Ultra-fast Vapor Generation by a Graphene Nano-ratchet

Hongru Ding[1,2,#], Guilong Peng[1,2,#], Dengke Ma[1,2], S.W. Sharshir[1,2] and Nuo Yang[1,2,*]

[1]State Key Laboratory of Coal Combustion, Huazhong University of Science and Technology (HUST), Wuhan 430074, P. R. China

[2]Nano Interface Center for Energy(NICE), School of Energy and Power Engineering, Huazhong University of Science and Technology (HUST), Wuhan 430074, P. R. China

[#] H. D. and G. P. contributed equally to this work.

Electronic mail: N.Y. (nuo@hust.edu.cn)

# SI. Molecular dynamic simulation details

| Method | Non- Equilibrium MD | | | | | |
|---|---|---|---|---|---|---|
| **Potential** | | | | | | |
| Function | $$E = \sum_{i}^{on\ a} \sum_{j}^{on\ b} \frac{k_c q_i q_j}{r_{ij}} + \frac{A}{r_{oo}^{12}} - \frac{B}{r_{oo}^{6}} + 4\epsilon[(\frac{\sigma}{r})^{12} - (\frac{\sigma}{r})^{6}]$$ | | | | | |
| **Parameters (TIP4P)** | $m_O$ | $m_H$ | $q_M$ (C) | $q_H$ (C) | $R_{OM}$ (Å) | $R_{coul,cut}$ (Å) |
| | 15.999 | 1.008 | -1.040 | 0.520 | 0.15 | 8.5 |
| | $R_{OH}$ (Å) | $\theta_{OH}$ (°) | $A \times 10^{-3}$ (Kcal Å$^{12}$/(mol)) | | B (Kcal Å$^6$/(mol)) | |
| | 0.9572 | 104.52 | 600.0 | | 610.0 | |
| **Parameters (VDW)** | Type of molecular | $N_2^*$ | $O_2^*$ | C | *Two centered LJ potential for $N_2$ and $O_2$ | |
| | $\epsilon$ (Kcal/(mol*Å$^2$)) | 0.0725 | 0.1034 | 0.05528 | | |
| | $\Sigma$ (Å) | 3.32 | 2.99 | 3.415 | | |
| **Simulation process** | | | | | | |
| Ensemble | Setting | | | | Purpose | |
| NVT | Time step (fs) | 1 | Runtime (ns) | 3 | Relax structure | |
| | Temperature (K) | 300 | Thermostat | Nosé–Hoover | | |
| | Simulation cell (nm) | 13.6*12.2*22 | | | | |
| | Boundary condition | X, Y, Z: periodic, periodic, fixed | | | | |
| NVT | Runtime (ns) | 8 | Temperature (K) | 300 | Record information | |
| | Simulation cell (nm) | 13.6*12.2*22 | | | | |
| | Boundary condition | X, Y, Z: periodic, periodic, fixed | | | | |

# SII. Test of timestep

In molecular dynamic simulations, the timestep size is constrained by the demand that the vibrational motion of the atoms be accurately tracked. Usually, timestep is limited to femtosecond scale[36]. To perform accurate but economic simulations, we made a test about different timesteps. We fix $N_l$, $A_{min}$ and $\tan\alpha$ as 3, 33.1Å$^2$ and 0.25, in this part. We calculate the pressure difference of MGCN with different timesteps at room temperature, the results are shown in Fig S1. The values of $\Delta P$ change little when the timesteps are smaller than 1 fs. It shows that our simulation cell is large enough to overcome the finite size effect on calculating thermal conductivity. In all of the simulations of MGCN, we use 1 fs as the timestep.

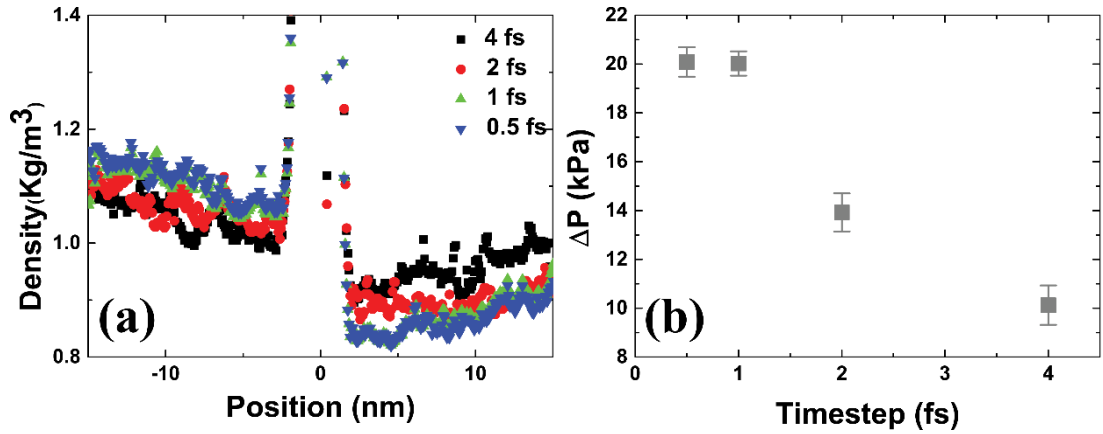

Figure S1. The density profile of saturated moist air molecules (a) and the pressure difference (b) with different timesteps. In this simulation, MGCN is constructed by 3 layers graphene with truncated cone nanopores. The area of minimum pore is 33.1 Å$^2$ and the cone angle $\alpha$ is arctan0.25.

## SIII. Finite size effect

In this simulation, the finite size effect could arise if the simulation cell is not sufficiently large. As shown in Figure S2, we calculate the pressure difference $\Delta P$ of MGCN with different sizes at room temperature. We fix the number of graphene layers

$N_l$, the area of the minimum pore $A_{min}$ and the cone angle α as 2, 33.1 Å² and arctan0.25, respectively. The values of ΔP change little when the size of simulation cell is larger than 4×4 unit cells (UCs). It shows that our simulation cell is large enough to overcome the finite size effect on calculating ratchet transport. In all of the simulations of MGCN, we use 4×6 UCs as simulation cell.

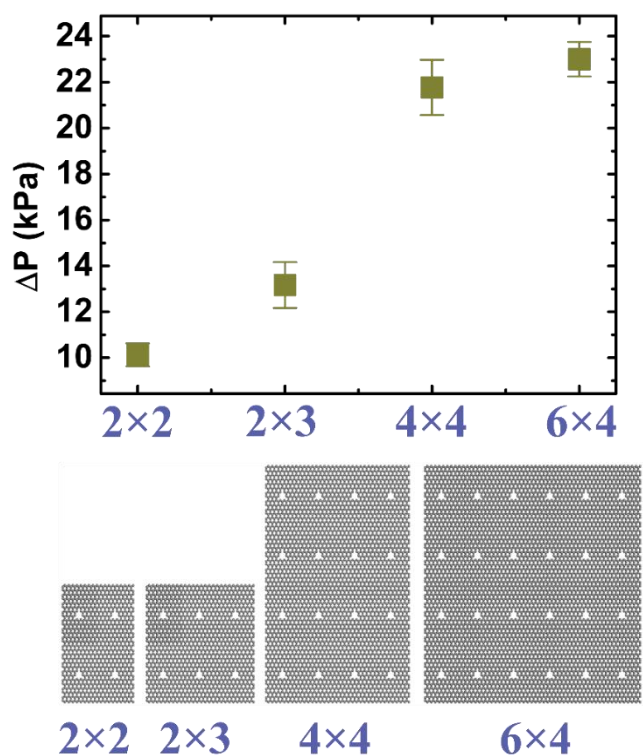

Figure S2. The pressure difference versus simulation cells. In this simulation, MGCN is constructed by 2 graphene layers with cone-shaped nanopores, where the area of minimum pore is 33.1 Å² and the cone angle α is arctan0.25..

## SIV. Dependence of ambient temperature

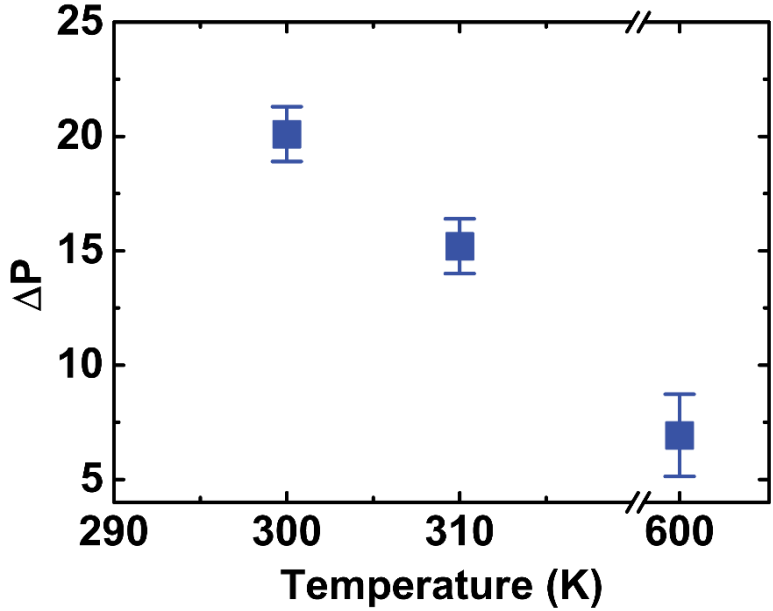

Figure S3. The pressure difference versus ambient temperature. ΔP shows a significant reduction, when ambient temperature increase from 300 K to 600 K. In this simulation, MGCN is constructed by 4 graphene layers with cone-shaped nanopores, where the area of minimum pore is 55.2 Å² and the cone angle α is arctan0.25.

## SV. Dependence of truncated-cone angle α

ΔP also has nonmonotonic dependence on the truncated-cone angle α. In the following cases, $N_l$ and $A_{min}$ are fixed as 3 and 33.1 Å², respectively. As shown in Fig. 2(d), we get the biggest ΔP, 20.1 kPa, when tanα is 0.25. Figure S4 shows the force distribution of MGCN with different α. As for the reasons of the low ΔP for the other three α: on the one hand, the narrow nanochannel and many energy barriers of the small-α MGCN confine the ratchet effect; On the other hand, the strong diffusion transport and weak negative force of the big-α MGCN result in the low ΔP. The discussions about diffusion transport are described in details below. That is why the adopted tanα of other simulations is fixed at 0.25 in this paper.

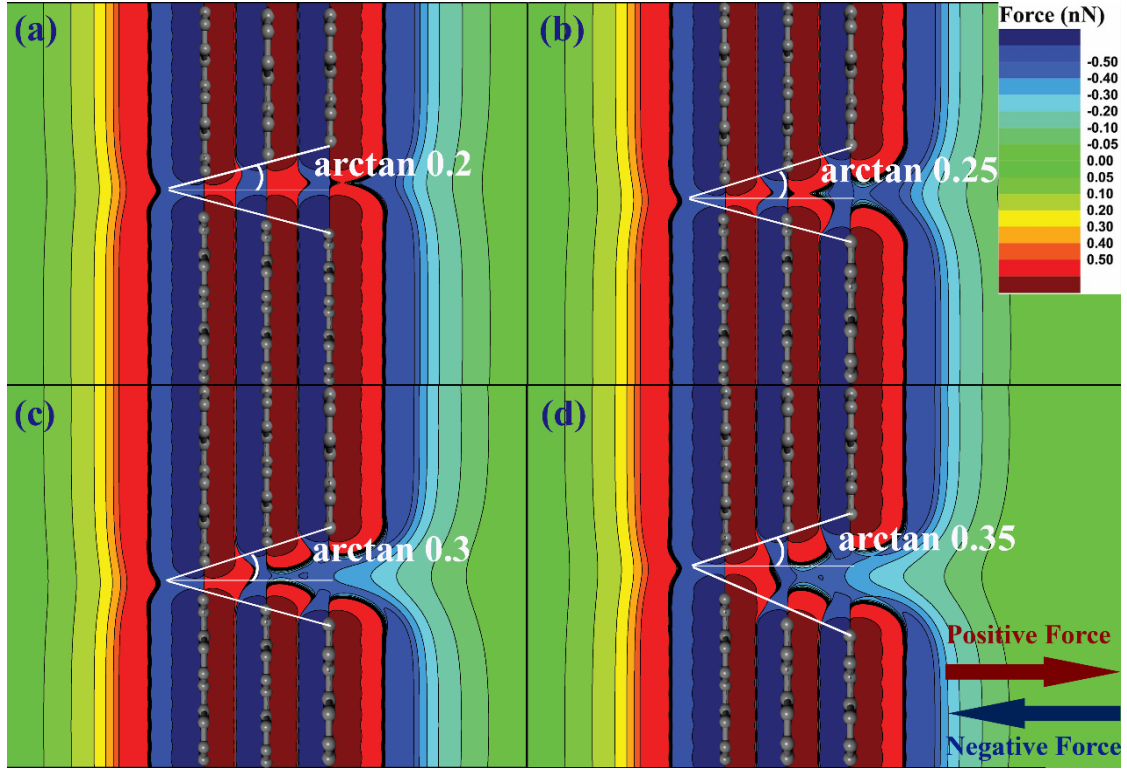

Figure S4. The number of graphene layers is fixed as 3; and the area of minimum pore $A_m$ as 33.1 Å$^2$. The force (along the Z direction) distribution of MGCN with different cone angle, arctan0.2- arctan0.35, are shown in (a)-(d). The arrows point out the direction of positive and negative force.

## SVI. Knudsen Diffusion

Diffusive transport, resulted from the concentration difference, also affects the molecular transport. The molecules will move from the high concentration region to the low due to the diffusive transport, and this limits the further growth of concentration. Therefore, the final distribution of air depends on the competition between the ratchet transport (leftward) and diffusive transport (rightward). Since the scale length of the truncated-cone nanopores is much smaller than the mean free path of the air molecules, the Knudsen diffusion[42] occurs here. The Knudsen diffusion flux is defined as,

$$\Phi_k = -\frac{2}{3}\bar{r}u\frac{dC}{dz} \tag{S1}$$

$$\bar{r} = (r_{min} + r_{max})/2 \tag{S2}$$

$$r_{max} = r_{min} + (N_l - 1)h\tan\alpha \tag{S3}$$

where $\bar{r}$ is the mean radium of the nanochannel, $r_{min}$ and $r_{max}$ are radiuses of the minimum and maximum pores. $r_{min}$ equals to $\sqrt{A_{min}/\pi}$; $u$ is the characteristic velocity of air molecules; $\frac{dC}{dz}$ is the concentration gradient and $h$ is the interlayer spacing of MGCN. In equation (S1), $\bar{r}$ and $\Phi_k$ are substituted by (S2)-(S3). Then equation (S1) can be defined as,

$$\Phi_K = -\frac{2r_{min} + (N_l - 1)h\tan\alpha}{3} u \frac{dC}{dz} \tag{S4}$$

## SVII. Evaporation enhancement calculation

According to Hertz-Knudsen Relation[44] as defined here:

$$\dot{m} = (\sigma_e \frac{P_S}{\sqrt{T_L}} - \sigma_c \frac{P_V}{\sqrt{T_a}}) \sqrt{\frac{M}{2\pi R}} \tag{S5}$$

where $\dot{m}$ is the evaporation rate of the water, $P_S$ and $P_V$ are the water vapor saturate pressure and the real vapor partial pressure at the interface respectively. $\sigma_e$ and $\sigma_c$ are the evaporation and condensation coefficient, respectively. $M$ is the molar mass of the water molecule. $T_L$ and $T_a$ are the temperature of the water and vapor at the interface respectively.

Normally, $\sigma_e$ and $\sigma_c$ are measured at the range of 0.001 to 1 and very close to each other [44-46]. $T_L$ is slightly higher than $T_a$ when water is heated to evaporate.[43] And $P_V$ is lower than $P_S$ due to the lower vapor temperature and molecular diffusion. Therefore, the following assumptions are made: (i) The temperature discontinuity at the water-vapor interface is ignored, i.e., $T_L = T_a = T$; (ii) The difference between evaporation and condensation coefficient is ignored, i.e., $\sigma_e = \sigma_c = \varepsilon$.

when MGCN is applied, the enhancement of evaporation, $\eta_i$, can be calculated as:

$$\eta_i = \frac{P_V - P_V'}{(P_S - P_V)} \times 100\% \tag{S6}$$

where $P_V'$ and $P_V$ are the real vapor saturate pressure at the interface with and without MGCN respectively. According to equation (1), $P_V$ can be described as:

$$P_V = P_S - \frac{\dot{m}}{\varepsilon}\sqrt{\frac{2\pi RT}{M}} \qquad (S7)$$

Meanwhile, as shown in Figure S4, due to the diffusion resistance, the vapor would accumulate on the high pressure side, which indicates that the pressure on the high pressure side of MGCN can be regarded as $P_S$. Therefore, $P'_V$ can be determined by:

$$P'_V = \left(1 - \frac{\Delta P}{P_{atm}}\right) P_S \qquad (S8)$$

where $P_{atm}$ is the atmospheric pressure, $\Delta P$ is the pressure difference between the two sides of MGCN. Hence, $\eta_i$ can be described as:

$$\eta_i = \left(\frac{\varepsilon \cdot \Delta P \cdot P_S}{P_{atm}\dot{m}}\sqrt{\frac{M}{2\pi RT}} - 1\right) \times 100\% \qquad (S9)$$

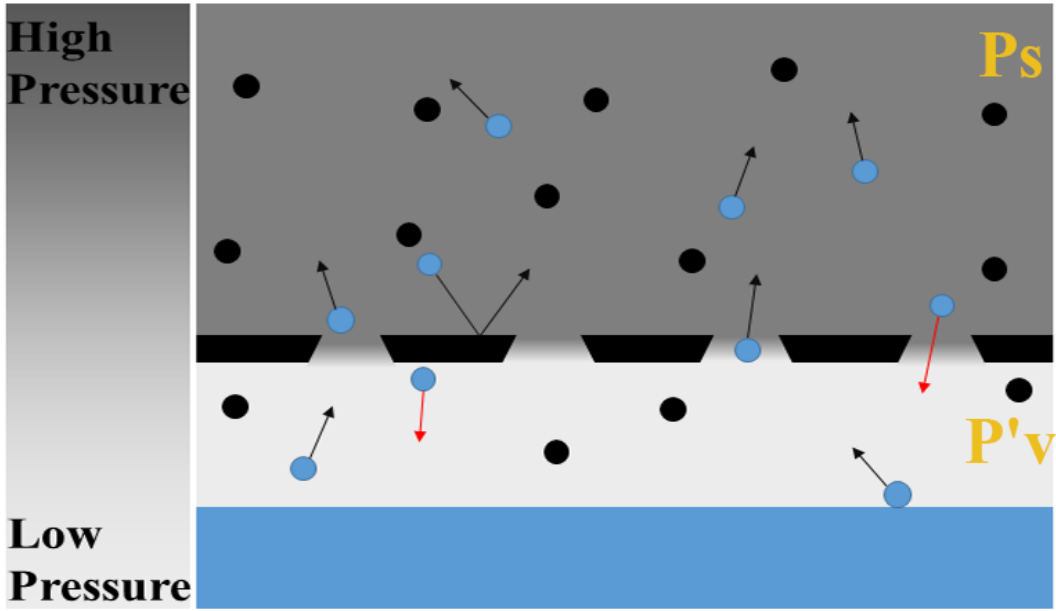

Figure S5    Vapor pressure on both sides of MGCN. The pressure at the interface, $P'_V$, is low due to the pumping by MGCN, the pressure on the other side is $P_S$ due to the accumulation of vapor molecules.

**5** 4449 (2014).